\documentclass[showpacs,aps,epsfig,nofootinbib]{revtex4}
\usepackage{graphicx}
\usepackage{epstopdf}
\usepackage{epsfig}
\usepackage{latexsym}
\usepackage{amssymb}
\usepackage{color}
\usepackage[center]{subfigure}

\begin{document}

%%%%%%%%%%%%%%%%%%%%%%%%%%%%%%%%%%%%%%%%%%%%%%%%%%%%%%%%%%%%%%%
 \newcommand{\bq}{\begin{equation}}
 \newcommand{\eq}{\end{equation}}
 \newcommand{\bqn}{\begin{eqnarray}}
 \newcommand{\eqn}{\end{eqnarray}}
 \newcommand{\nb}{\nonumber}
 \newcommand{\lb}{\label}
\newcommand{\PRL}{Phys. Rev. Lett.}
\newcommand{\PL}{Phys. Lett.}
\newcommand{\PR}{Phys. Rev.}
\newcommand{\CQG}{Class. Quantum Grav.}
 %%%%%%%%%%%%%%%%%%%%%%%%%%%%%%%%%%%%%%%%%%%%%%%%%%%%%%%%%%%%%%%

\title[Fermions tunneling from higher-dimensional Reissner-Nordstr\"{o}m black hole]{Fermions tunneling from higher-dimensional Reissner-Nordstr\"{o}m black hole: semiclassical and beyond semiclassical approximation}

\author{ShuZheng Yang$^{1)}$}
\email{szyangcwnu@126.com}
\author{Dan Wen$^{1,2,3)}$}
\email{danwen111@163.com}
\author{Kai Lin$^{1,4)}$}
\email{lk314159@hotmail.com}

\affiliation{$^{1)}$Department of Astronomy, China West Normal
University, Nanchong, Sichuan 637002 China}

\affiliation{$^{2)}$College of Physics and Information Science,
Hunan Normal University, Changsha, Hunan 410081, China}

\affiliation{$^{3)}$Faculdade de Engenharia de Guaratinguet\'a,
Universidade Estadual Paulista, Guaratinguet\'a, SP, Brasil}

\affiliation{$^{4)}$Instituto de F\'isica e Qu\'imica, Universidade
Federal de Itajub\'a,Itajub\'a, MG, Brasil}

\begin{abstract}

Based on semiclassical tunneling method, we focus on charged
fermions tunneling from higher-dimensional Reissner-Nordstr\"{o}m
black hole. We first simplify the Dirac equation by semiclassical
approximation, and then a semiclassical Hamilton-Jacobi equation is
obtained. Using the Hamilton-Jacobi equation, we study the Hawking
temperature and fermions tunneling rate at the event horizon of the
higher-dimensional Reissner-Nordstr\"{o}m black hole spacetime.
Finally, the correct entropy is calculation by the method beyond
semiclassical approximation.



\end{abstract}

%Uncomment for PACS numbers title message
%\pacs{00.00, 20.00, 42.10}
% Keywords required only for MST, PB, PMB, PM, JOA, JOB?
%\vspace{2pc}
%\noindent{\it Keywords}: Article preparation, IOP journals
% Uncomment for Submitted to journal title message
%\submitto{\JPA}
% Comment out if separate title page not required
\maketitle Hawking radiation is an important prediction in modern
gravitation theory \cite{1,2,3,4,5}. Recently, Kraus, Parikh and
Wilczek proposed quantum tunneling theory to explain and study
Hawking radiation
\cite{6,7,8,9,10,11,12,13,14,15,16,17,18,19,20,21,22,23,24,25,26,27,28,29,30,31},
and then Semiclassical Hamilton-Jacobi method is put forward to
research the properties of scalar particles' tunnels
\cite{32,33,34,35,36}. In 2007, Kerner and Mann investigated the 1/2
spin fermion tunneling from static black holes \cite{37}. In their
work, the spin up and spin down cases are researched respectively,
and the radial equations are obtained, so that they can finally
determine the Hawking temperature and tunneling rate at the event
horizon. Subsequently, Kerr and Kerr-Newman black holes cases, the
charged dilatonic black hole case, the de Sitter horizon case, the
BTZ black hole case, 5-dimensional spacetime cases and several
non-stationary black hole cases were all researched respectively
\cite{38,39,40,41,42,43,44,45,46,47,48}, and we used Hamilton-Jacobi
method to study the fermion tunneling from higher-dimensional
uncharged black holes \cite{49,50}. However, up to now, no one has
studied higher-dimensional charged black holes cases, so we set out
to research that case. In our work, we developed the Kerner and Mann
method, and proved the semiclassical Hamilton-Jacobi equation not
only can be obtained with the Klein-Gordon equation of curved
spacetime, but also with the Dirac equation in curved spacetime.
Applying the Hamilton-Jacobi equation, we can then obtain
semiclassical Hawking temperature and tunneling rate at the event
horizon of higher-dimensional Reissner-Nordstr\"{o}m black hole.

In modern physics theory, the concept of an extra dimension can help
to solve some theoretical issues, so several higher dimensional
metrics of curved spacetime was investigated. The metric of static
charged $(n+2)$-dimensional Reissner-Nordstr\"{o}m black hole is
given by \cite{10, 51,52,53,54,55}
\begin{equation}
\label{eq1}ds^2=-f(r)dt^2+f^{-1}(r)dr^2+r^2d\Omega^2_n
\end{equation}
where $d\Omega^2_n$ is the metric of $n$-dimensional sphere
\begin{equation}
\label{eq2}d\Omega^2_n=\sum_{i=1}^{n}h^{ii}d\theta^2_i=d\theta^2_1+\sin^2\theta_1d\theta^2_2+\sin^2\theta_1\sin^2\theta_2d\theta^2_3+\cdot\cdot\cdot+\prod^{n-1}_{i=1}\sin^2\theta_id\theta^2_n
\end{equation}
and
\begin{equation}
\label{eq3}f(r)=1-\frac{\omega_nM}{r^{n-1}}+\frac{\omega_nQ^2}{2(n-1)V_nr^{2n-2}}~~~~~~~~\omega_n=\frac{16\pi}{nV_n}
\end{equation}
$M$ and $Q$ are mass and electric charge of black hole, and the
electro-magnetic potential is
\begin{equation}
\label{eq4}A_\mu=\left(\frac{Q}{(n-1)V_nr^{n-1}},0,0,0,\cdot\cdot\cdot\right)
\end{equation}
where $V_n$ is volume of unit n-sphere (we can adopt the units
$G=c=\hbar=1$). The outer/ inner horizon located at
\begin{equation}
\label{eq5}r^{n-2}_\pm=\frac{\omega_n}{2}\left[M\pm
\sqrt{M^2-\frac{nQ^2}{8\pi (n-1)}} \right]
\end{equation}
Obviously, at the horizons, the equation $f(r_\pm)=0$ should be
satisfied. However, the physical property near the inner horizon
cannot be researched, so we just study the fermion tunneling at the
outer event horizon of this black hole. The charged Dirac equation
in curved spacetime is
\begin{equation}
\label{eq6}\gamma^\mu D_\mu
\Psi+\frac{m}{\hbar}\Psi=0~~~~~~~\mu=t,r,\theta_1,\cdot\cdot\cdot,\theta_n
\end{equation}
where
\begin{equation}
\label{eq7}D_\mu=\partial_\mu+\Gamma_\mu+\frac{iqA_\mu}{\hbar}
\end{equation}
\begin{equation}
\label{eq8}\Gamma_\mu=\frac{1}{8}\left[\tilde\gamma^a,\tilde\gamma^b\right]e_a^{~\nu}e_{b\nu;\mu}
\end{equation}
$m$ and $q$ are mass and electric charge of the particles, and
$e_{b\nu;\mu}=\partial_\mu e_{b\nu}-\Gamma^\alpha_{\mu\nu}e_{ab}$ is
the covariant derivative of tetrad $e_{b\nu}$. The gamma matrices in
curved spacetime need to be satisfied
\begin{equation}
\label{eq9}\{\gamma^\mu,\gamma^\nu\}=2g^{\mu\nu}I
\end{equation}
After the gamma matrices are defined, we choose the gamma matrices
in (n+2)-dimensional flat spacetime as
\begin{equation}
\label{eq10} \tilde {\gamma}^1_{m\times
m}=\left(\begin{array}{cc}I_{\frac{m}{2}\times \frac{m}{2}} & 0 \\
0& -I_{\frac{m}{2}\times \frac{m}{2}}
\end{array}\right)
\end{equation}
\begin{equation}
\label{eq11} \tilde {\gamma}^2_{m\times
m}=\left(\begin{array}{cc} 0 & I_{\frac{m}{2}\times \frac{m}{2}}\\
I_{\frac{m}{2}\times \frac{m}{2}} & 0
\end{array}\right)
\end{equation}
\begin{equation}
\label{eq12} \tilde {\gamma}^\eta_{m\times
m}=\left(\begin{array}{cc} 0 & i\tilde{\gamma}^{\eta-2}_{\frac{m}{2}\times \frac{m}{2}}\\
-i\tilde{\gamma}^{\eta-2}_{\frac{m}{2}\times \frac{m}{2}} & 0
\end{array}\right)~~~~~~~~~~~~~~~~\eta=3,4,5,\cdot\cdot\cdot,n+2
\end{equation}
where $I_{\frac{m}{2}\times \frac{m}{2}}$ and
$\tilde{\gamma}^\eta_{\frac{m}{2}\times \frac{m}{2}}$ are unit
matrices and flat gamma matrices with $\frac{m}{2}\times
\frac{m}{2}$ order, and $m=2^{(n+2)/2}$ is the order of the matrices
in even (odd) dimensional space-time. Corresponding to the flat
case, the gamma matrices can be chosen as
\begin{equation}
\label{eq13} \gamma^t_{m\times m}=\frac{i}{\sqrt{f}}\tilde
{\gamma}^1_{m\times m}
\end{equation}
\begin{equation}
\label{eq14} \gamma^r_{m\times m}=\sqrt{f}\tilde {\gamma}^2_{m\times
m}
\end{equation}
\begin{equation}
\label{eq15} \gamma^\eta_{m\times m}=r^{-1}\sqrt{h^{\eta\eta}}\tilde
{\gamma}^\eta_{m\times m}~~~~~~~~~\eta=3,4,5,\cdot\cdot\cdot,n+2.
\end{equation}

Now, let's simplify the Dirac equation via semiclassical
approximation, and rewrite the spinor function as
\begin{equation}
\label{eq16} \Psi=\left(\begin{array}{c} A_{\frac{m}{2}\times 1}(t,r,\cdot\cdot\cdot, x^\eta,\cdot\cdot\cdot)\\
B_{\frac{m}{2}\times 1}(t,r,\cdot\cdot\cdot, x^\eta,\cdot\cdot\cdot)
\end{array}\right)e^{\frac{i}{\hbar}S(t,r,\cdot\cdot\cdot,x^\eta,\cdot\cdot\cdot)}
\end{equation}
where $A_{\frac{m}{2}\times 1}(t,r,\cdot\cdot\cdot
x^\eta,\cdot\cdot\cdot)$ and $B_{\frac{m}{2}\times
1}(t,r,\cdot\cdot\cdot, x^\eta,\cdot\cdot\cdot)$ are are column
matrices with $\frac{m}{2}\times 1$ order, and S is classical
action. Via semiclassical approximation method, Substituting Eq.(16)
into Eq.(6) and dividing the exponential term and multiplying by
$\hbar$, we can get
\begin{equation}
\label{eq17}\left(\begin{array}{cc} C & D\\
E & F
\end{array}\right)\left(\begin{array}{c} A_{\frac{m}{2}\times 1}\\
B_{\frac{m}{2}\times 1}
\end{array}\right)=0
\end{equation}
\begin{equation}
\label{eq18} C=-\frac{1}{\sqrt{f}}\left(\frac{\partial S}{\partial
t}+qA_t\right)I_{\frac{m}{2}\times
\frac{m}{2}}+mI_{\frac{m}{2}\times \frac{m}{2}}
\end{equation}
\begin{equation}
\label{eq19} D=i\sqrt{f}\frac{\partial S}{\partial
r}I_{\frac{m}{2}\times \frac{m}{2}}-\sum_\eta
r^{-1}\sqrt{h^{\eta\eta}}\frac{\partial S}{\partial x^\eta} \tilde
{\gamma}^{\eta-2}_{\frac{m}{2}\times \frac{m}{2}}
\end{equation}
\begin{equation}
\label{eq20} E=i\sqrt{f}\frac{\partial S}{\partial
r}I_{\frac{m}{2}\times \frac{m}{2}}+\sum_\eta
r^{-1}\sqrt{h^{\eta\eta}}\frac{\partial S}{\partial x^\eta} \tilde
{\gamma}^{\eta-2}_{\frac{m}{2}\times \frac{m}{2}}
\end{equation}
\begin{equation}
\label{eq21} F=\frac{1}{\sqrt{f}}\left(\frac{\partial S}{\partial
t}+qA_t\right)I_{\frac{m}{2}\times
\frac{m}{2}}+mI_{\frac{m}{2}\times \frac{m}{2}}
\end{equation}
Solving Eq.(17), we have
\begin{equation}
\label{eq22} (E-FD^{-1}C)A_{\frac{m}{2}\times 1}=0
\end{equation}
\begin{equation}
\label{eq23} (F-EC^{-1}D)B_{\frac{m}{2}\times 1}=0
\end{equation}
It is evident that the coefficient matrices of (22) and (23) must
vanish, when $A_{\frac{m}{2}\times 1}$ and $B_{\frac{m}{2}\times 1}$
have non-trivial solutions. Due to the fact that $CD=DC$, we can
write the condition that determinant of coefficient vanish as
\begin{equation}
\label{eq24} \text{det}(ED-FC)=0
\end{equation}
From the relation of flat gamma matrices $\{\tilde
{\gamma}^\mu,\tilde {\gamma}^\nu\}=2\delta_{\mu\nu}$, we can obtain
the semiclassical Hamilton-Jacobi equation in $(n+2)$-dimensional
Reissner-Nordstr\"{o}m space time
\begin{equation}
\label{eq25}-\frac{1}{f}\left(\frac{\partial S}{\partial
t}+qA_t\right)^2+f\left(\frac{\partial S}{\partial
r}\right)^2+\cdot\cdot\cdot+g^{\eta\eta}\left(\frac{\partial
S}{\partial x^\eta}\right)+\cdot\cdot\cdot+m^2=0
\end{equation}
Using the Hamilton-Jacobi equation, in charged static spacetime, we
can separate the variables for the action as
\begin{equation}
\label{eq26}S=-\omega
t+R(r)+Y(\cdot\cdot\cdot,x^\eta,\cdot\cdot\cdot)+K~~~~~(K~\text{is a
constant})
\end{equation}
and the Hamilton-Jacobi equation is broken up as
\begin{equation}
\label{eq27}-\frac{1}{f}\left(\omega-qA_t\right)^2+f\left(\frac{dR}{dr}\right)^2+m^2=\frac{\lambda}{r^2}
\end{equation}
\begin{equation}
\label{eq28} \sum_\eta h^{\eta\eta}\left(\frac{\partial Y}{\partial
x^\eta}\right)^2+\lambda=0
\end{equation}
where Eq.(27) and Eq.(28) are radial and non-radial equations
respectively, and $\lambda$ is a constant. However, we only research
on the radial equation, because the tunnel at the event horizon is
radial. From Eq.(27), we can get
\begin{equation}
\label{eq29}
\frac{dR(r)}{dr}=\pm\frac{\sqrt{(\omega-qA_t)^2r^2+f(\lambda-m^2r^2)}}{fr}
\end{equation}
Near the event horizon, the radial action is given by
\begin{equation}
\label{eq30} \text{Im} R_\pm=\pm\frac{\pi
(\omega-\omega_0)}{f^\prime (r_+)}+\text{Im} C
\end{equation}
where, $R_+$ is part of the outgoing solution, while $R_-$ is the
part of incoming solution, and
\begin{equation}
\label{eq31} \omega_0=q\frac{Q}{(n-1)V_nr^{n-1}_+}
\end{equation}
So the tunneling rate is
\begin{equation}
\label{eq32}
\Gamma=\frac{\text{Prob[out]}}{\text{Prob[in]}}=\frac{\text{exp}(-2\text{Im}S_+)}{\text{exp}(-2\text{Im}S_-)}=\frac{\text{exp}(-2\text{Im}R_++\text{Im}K)}{\text{exp}(-2\text{Im}R_-+\text{Im}K)}=\text{exp}\left(\frac{-4\pi
(\omega-\omega_0)}{f^\prime (r_+)}\right)
\end{equation}
where $\text{Im}$ represents the imaginary part of the function, and
the Hawking temperature is
\begin{equation}
\label{eq33} T_H=\frac{f^\prime (r_+)}{4\pi}
\end{equation}

However, above calculation is worked on semiclassical approximation,
because we ignored all higher order terms of ${\cal O}(\hbar)$.
Recently, Banerijee and Majhi proposed a new method beyond
semiclassical approximation to research the quantum tunneling, and
their works show that the conclusion should be corrected
\cite{56,57,58,59,60,61,62,63,64,65}, and this correct entropy may
be applied in quantum gravity theory. Now let's generalize this work
in higher-dimensional Reissner-Nordstr\"{o}m black hole spacetime.

Because the tetrad $e_\mu^{~a}$ in the spacetime are given by
\begin{equation}
\label{eq34}
e_\mu^{~a}=\text{diag}\left(\sqrt{f},\frac{1}{\sqrt{f}},r,r\sin\theta_1,\cdot\cdot\cdot,r\prod_{i=1}^{n-1}\sin\theta_i\right),
\end{equation}
so that $\Gamma_\mu$ is
\begin{equation}
\label{eq35} \tilde\gamma^a
e^{~\mu}_a\Gamma_\mu=\tilde\gamma^1\sqrt{f}\left(\frac{n}{2r}+\frac{f'}{4f}\right)+\frac{1}{2r}\sum_{k=1}^{n-1}\tilde\gamma^{k+1}\frac{(n-k)\cot\theta_k}{\prod_{i=1}^{k-1}\sin\theta_i}.
\end{equation}
It means the Dirac equation becomes
\begin{eqnarray}
\label{eq36}
i\frac{\tilde\gamma^0}{\sqrt{f}}\left(\frac{\partial}{\partial
t}+\frac{iqA_t}{\hbar}\right)\Psi+\tilde\gamma^1\sqrt{f}\left(\frac{\partial}{\partial
r}+\frac{n}{2r}+\frac{f'}{4f}\right)\Psi&&\nonumber\\
+\sum_{k=1}^{n-1}\frac{\tilde\gamma^{k+1}}{r\prod_{i=1}^{k-1}\sin\theta_i}\left(\frac{\partial}{\partial
\theta_k}+\frac{(n-k)\cot\theta_k}{2}\right)\Psi+\frac{\tilde\gamma^{n+1}}{r\prod_{i=1}^{n-1}\sin\theta_i}\frac{\partial
\Psi}{\partial\theta_n}+\frac{m}{\hbar}\Psi&=&0,
\end{eqnarray}
and this equation can be simplified at event horizon
\begin{equation}
\label{eq37} i\tilde\gamma^0\left(\frac{\partial}{\partial
t}+\frac{iqA_0}{\hbar}\right)\Psi+\tilde\gamma^1\left(\frac{\partial}{\partial
r_*}+\frac{f'}{4}\right)\Psi=0,
\end{equation}
because $f\rightarrow0$ at event horizon, and $dr_*=\frac{dr}{f}$ is
tortoise coordinate. On the other hand, the spacetime background is
static, and $\Psi$ can be rewritten as
\begin{equation}
\label{eq38}\Psi=\left[\begin{array}{c} A(r)\\
B(r)
\end{array}\right]e^{-\frac{i}{\hbar}\omega t},
\end{equation}
where $A(r)$ and $B(r)$ are matrices with $\frac{m}{2}\times1$, and
$\omega$ is frequency or energy of Dirac particle. Finally, apply
the definitions of $\tilde\gamma^0$ and $\tilde\gamma^1$, we get
\begin{equation}
\label{eq39}\left(\begin{array}{cc} \omega-\omega_0 & \hbar f\left(\frac{\partial}{\partial r}-\frac{f'}{4f}\right)\\
\hbar f\left(\frac{\partial}{\partial r}-\frac{f'}{4f}\right) &
-(\omega-\omega_0)
\end{array}\right)\left(\begin{array}{c} A_q\\
B_q
\end{array}\right)=0,
\end{equation}
here $A_q$ and $B_q$ are $q$-th elements of matrices $A(r)$ and
$B(r)$ respectively. Above equation becomes
\begin{equation}
\label{eq40}\frac{\frac{\partial B_q}{\partial r}}{\frac{\partial
A_q}{\partial
r}}=\frac{(\omega-\omega_0)A_q-\hbar\frac{f'B_q}{4}}{-(\omega-\omega_0)B_q-\hbar\frac{f'A_q}{4}},
\end{equation}
so
\begin{equation}
\label{eq41}\frac{\omega-\omega_0}{2}\frac{\partial}{\partial
r}\left(A_q^2+B_q^2\right)-\hbar\frac{f'}{4}\left(B_q\frac{\partial
A_q}{\partial r}-A_q\frac{\partial B_q}{\partial r}\right)=0.
\end{equation}
At event horizon $f'(r_0)\not=0$ and depends on the position $r_0$,
so above equation implies
\begin{eqnarray}
\label{eq42} \frac{\partial}{\partial
r}\left(A_q^2+B_q^2\right)&=&0,\nonumber\\
B_q\frac{\partial A_q}{\partial r}-A_q\frac{\partial B_q}{\partial
r}&=&0,
\end{eqnarray}
and the solution is
\begin{eqnarray}
\label{eq43} A_q^2+B_q^2=0.
\end{eqnarray}
Above relation means $A_q$ and $B_q$ can be rewritten as
\begin{eqnarray}
\label{eq44} A_q&=&C_qe^{\frac{i}{\hbar}R_q(r)},\nonumber\\
 B_q&=&F_qe^{\frac{i}{\hbar}R_q(r)},
\end{eqnarray}
and $C_q=\pm iF_q$ are constants.

Next, let's use the method beyond semiclassical approximation to
expand $R_q(r)$ and $K=\omega-\omega_0$ as
\begin{eqnarray}
\label{eq45} R_q&=&R_{q0}+\sum_{i=1}^{\infty}\hbar^iR_{qi}(r),\nonumber\\
 K&=&\omega-\omega_0=K_0+\sum_{i=1}^{\infty}\hbar^iK_i,
\end{eqnarray}
so we get
\begin{eqnarray}
\label{eq46}
\hbar^0:~~~\left(\begin{array}{cc} -i\frac{K_0}{f} & \frac{\partial R_{q0}}{\partial r}\\
\frac{\partial R_{q0}}{\partial r} & -i\frac{K_0}{f}
\end{array}\right)\left(\begin{array}{c} C_q\\
F_q
\end{array}\right)&=&0,\nonumber\\
\hbar^1:~~~\left(\begin{array}{cc} -i\frac{K_1}{f} & \frac{\partial R_{q1}}{\partial r}+\frac{if'}{4f}\\
\frac{\partial R_{q1}}{\partial r}+\frac{if'}{4f} & -i\frac{K_1}{f}
\end{array}\right)\left(\begin{array}{c} C_q\\
F_q
\end{array}\right)&=&0,\nonumber\\
\hbar^k:~~~\left(\begin{array}{cc} -i\frac{K_k}{f} & \frac{\partial R_{qk}}{\partial r}\\
\frac{\partial R_{qk}}{\partial r}& -i\frac{K_k}{f}
\end{array}\right)\left(\begin{array}{c} C_q\\
F_q
\end{array}\right)&=&0,~~~~~~k\ge2.
\end{eqnarray}
Therefore, the determinants of matrices vanish:
\begin{eqnarray}
\label{eq46}
\hbar^0:~~~R_{q0\pm}&=&\pm\int\frac{K_0}{f}dr,\nonumber\\
\hbar^1:~~~R_{q1\pm}&=&\pm\int\frac{K_1-\frac{if'(r_+)}{4}}{f}dr,\nonumber\\
\hbar^k:~~~R_{qk\pm}&=&\pm\int\frac{K_k}{f}dr,~~~~~~k\ge2,
\end{eqnarray}
and
\begin{eqnarray}
\label{eq47}
\text{Im}R_{qi}=\text{Im}R_{qi+}-\text{Im}R_{qi-}=\frac{2\pi
K_i}{f'(r_+)}.
\end{eqnarray}
In order to calculate the tunneling rate and Hawking temperature, we
rewrite $\text{Im}R_{qi}=\frac{\beta_i}{A_h^i}\text{Im}R_{q0}$
($i\ge0$, $A_h$ is area of black hole, and $\beta_i$ are
dimensionaless constant parameters) since the forms of $R_{qj}$ are
the same, so the total radial action is
\begin{eqnarray}
\label{eq48}
\text{Im}R_{q}=\text{Im}R_{q0}(r)+\sum_{i=1}^{\infty}\hbar^iR_{qi}(r)=\left(1+\sum_{i=1}^{\infty}\beta_i\frac{\hbar^i}{A_h^i}\right)R_{q0}.
\end{eqnarray}
the tunneling rate of Dirac particle at event horizon is given by
\begin{eqnarray}
\label{eq49}
\bar\Gamma_h=\text{exp}\left[-\frac{2}{\hbar}\left(1+\sum_{i=1}^{\infty}\beta_i\frac{\hbar^i}{A_h^i}\right)R_{q0}\right]=\text{exp}\left[-\frac{4\pi
K_0}{\hbar
f'(r_+)}\left(1+\sum_{i=1}^{\infty}\beta_i\frac{\hbar^i}{A_h^i}\right)\right].
\end{eqnarray}
From the relation between tunneling rate and Hawking radiation, we
get the Temperature of black holes
\begin{eqnarray}
\label{eq49}
T_h=\left(1+\sum_{i=1}^{\infty}\beta_i\frac{\hbar^i}{A_h^i}\right)T_H.
\end{eqnarray}
Finally, the laws of black hole thermodynamics requests
\begin{eqnarray}
\label{eq50} S_h=\int dS_h=\left.\int\frac{ dM-A
dQ}{T_h}\right|_{r=r_+}=\frac{A_h}{4\pi}+\pi
\beta_1\ln(A_h)+\cdot\cdot\cdot=S_H+\pi
\beta_1\ln(S_H)+\cdot\cdot\cdot
\end{eqnarray}
and $S_H=\frac{A_h}{4\pi}$ is entropy of semiclassical
approximation, and this result shows the correction of entropy is
logarithmic correction.

In this paper, we studied fermions tunneling from higher-dimensional
Reissner-Nordstr\"{o}m black holes, and obtained the Hamilton-Jacobi
equation from charged Dirac equation. This work shows that
semiclassical Hamilton-Jacobi equation can describe the property of
both 0 spin scalar particles and 1/2 spin fermions. In this work, we
did not emphasize dimensions of spacetime larger than (3+1)
dimensions, so the method also can be used in the research of
(3+1)-dimensions and lower cases.

As we all know, the information loss is an open problem in black
hole physics, and the information of particles maybe vanishes at the
singularity. In order to solve this difficulty, Horowitz and
Maldacena proposed a boundary condition, which is called as the
black hole final state, at singularity of black hole to perfectly
entangle between the incoming Hawking radiation particles and the
collapsing matter\cite{66,67}. Due to the boundary condition, any
particle which is falling into the black holes completely
annihilates. It is a new and interesting idea to investigate the
black hole physics and quantum gravity, so we also will work on this
area in the further.

\textbf{Acknowledgements:} This work is supported in part  by FAPESP
No. 2012/08934-0, CNPq, CAPES and National Natural Science
Foundation of China (No.11573022 and No.11375279)

\end{document}